\documentclass[sigconf]{acmart}

\usepackage{booktabs} 

\copyrightyear{2023}
\acmYear{2023}
\setcopyright{acmlicensed}
\acmConference[SAC '23]{The 37th ACM/SIGAPP Symposium on Applied Computing}{March 27-April 2, 2023}{Tallinn, Estonia} 
\acmBooktitle{The 37th ACM/SIGAPP Symposium on Applied Computing (SAC '23), March 27-April 2, 2023, Tallinn, Estonia}
\acmPrice{15.00}
\acmDOI{10.1145/3555776.3577682} \acmISBN{978-1-4503-9517-5/23/03}

\acmArticle{12927.99}

\usepackage{amsmath}
\usepackage{multirow}
\usepackage[linesnumbered,vlined,ruled]{algorithm2e}
\usepackage{pifont}
\newcommand{\cmark}{\ding{51}}%
\newcommand{\xmark}{\ding{55}}%

\DeclareMathOperator*{\argmax}{argmax}
\DeclareMathOperator*{\avg}{avg}

\begin{document}
\title{Diversity-Promoting Ensemble for Medical Image Segmentation}
  
\renewcommand{\shorttitle}{Diversity-Promoting Ensemble for Medical Image Segmentation}


\author{Mariana-Iuliana Georgescu}
\affiliation{%
  \institution{University of Bucharest}
  \country{Romania} 
}

\author{Radu Tudor Ionescu}
\authornote{Corresponding author.}
\affiliation{%
  \institution{University of Bucharest}
  \country{Romania} 
}
\email{raducu.ionescu@gmail.com}

\author{Andreea-Iuliana Miron}
\affiliation{%
  \institution{``Carol Davila'' University of Medicine and Pharmacy, Col\c{t}ea Hospital}
  \country{Romania}}


\begin{abstract}
Medical image segmentation is an actively studied task in medical imaging, where the precision of the annotations is of utter importance towards accurate diagnosis and treatment. In recent years, the task has been approached with various deep learning systems, among the most popular models being U-Net. In this work, we propose a novel strategy to generate ensembles of different architectures for medical image segmentation, by leveraging the diversity (decorrelation) of the models forming the ensemble. More specifically, we utilize the Dice score among model pairs to estimate the correlation between the outputs of the two models forming each pair. To promote diversity, we select models with low Dice scores among each other.
We carry out gastro-intestinal tract image segmentation experiments to compare our diversity-promoting ensemble (DiPE) with another strategy to create ensembles  based on selecting the top scoring U-Net models. Our empirical results show that DiPE surpasses both individual models as well as the ensemble creation strategy based on selecting the top scoring models.
\end{abstract}

%
%


\maketitle

\section{Introduction}

Physicians extensively use medical imaging techniques, e.g.~Computed Tomography (CT), Magnetic Resonance Imaging (MRI) and Optical Coherence Tomography (OCT) \cite{Verga-RRP-2014}, as one of the least invasive investigation alternatives to diagnose lesions inside the human body. Segmenting (delimiting) regions of interest, such as organs or tumors, is often required for precise diagnosis and treatment. For example, a precise segmentation of a malignant tumor can lead to an accurate calibration of the radiation dosage in radiotherapy \cite{Jin-MIA-2021,Wang-NPL-2018,Men-PMB-2018, Seo-TMI-2020}. In recent years, the medical image segmentation task has been approached with various deep learning systems, ranging from convolutional neural networks (CNNs) \cite{Ronneberger-MICCAI-2015,Seo-TMI-2020,Zahangir-JMI-2019} to transformers \cite{Chen-arXiv-2021,Gao-MICCAI-2021,Hatamizadeh-WACV-2022}. Among these, U-Net \cite{Ronneberger-MICCAI-2015} remains one of the most popular methods. Although U-Net was introduced in 2015, it consistently received updates \cite{Siddique-Access-2021,Zhou-DLMIAMLCDC-2018,Zahangir-JMI-2019,Oktay-Arxiv-2018}, keeping its performance at a competitive level. However, using a single neural network to perform segmentation is not always the best solution. Indeed, constructing ensembles of multiple neural networks is an extensively validated method \cite{Lyksborg-SCIA-2015,Baldeon-NN-2020,Zhao-MICCAI-2019,Ganaie-EAAI-2022} to boost accuracy. 

Since the precision of the medical image segmentation output is of utter importance towards accurate diagnosis and treatment, we focus on combining multiple U-Net architectures to address the task. We conjecture that decorrelated models lead to a superior ensemble, since decorrelated models can better complement each other's decisions. To this end, we propose a novel strategy to construct ensembles of different models for medical image segmentation by promoting the diversity (decorrelation) of the models comprising the ensemble, while also giving equal importance to accuracy. To measure the correlation among two models, we compute the Dice score between the outputs of the respective models. We then construct the ensemble in a bottom-up fashion, starting from the best model and gradually adding the least correlated models with those already included, one by one. At the same time, our ensemble creation strategy assigns equal importance to the performance level of the model to be added at each step. Since we select models with lower Dice scores at each step, our strategy promotes the diversity among the models comprising the ensemble, hence bearing the name \emph{Diversity-Promoting Ensemble} (DiPE).

We conduct image segmentation experiments on the gastro-intestinal tract data set provided by the UW-Madison Carbone Cancer Center \cite{Lee-Kaggle-2022}. We evaluate nine individual U-Net models based on three different backbones (ResNet-34 \cite{He-CVPR-2016}, EfficientNet-B0 \cite{Tan-ICML-2019}, EfficientNet-B1 \cite{Tan-ICML-2019}) with or without multi-head convolutional attention \cite{Georgescu-WACV-2023}. Along with the individual models, we evaluate two strategies to create voting-based ensembles, namely $(i)$ a baseline (conventional) strategy selecting the top scoring models and $(ii)$ our strategy promoting diversity among the selected models. The empirical results indicate that our strategy, DiPE, outperforms both individual models, as well as the baseline ensemble. 

In summary, our contribution is twofold:
\begin{itemize}
    \item We introduce a diversity-promoting strategy to create an ensemble of medical image segmentation models that are low correlated among each other, by leveraging the Dice score between the outputs of various models.
    \item We provide empirical evidence showing that our diversity-promoting ensemble leads to superior performance levels compared with individual models and the conventional strategy selecting the top scoring models. 
\end{itemize}

\section{Related Work}

Medical image segmentation can be divided into two tasks, with respect to the input image. Indeed, there are works that tackle the segmentation task on 2D images \cite{Ronneberger-MICCAI-2015,Luo-MIA-2021,Zhao-TMI-2022,Wang-NPL-2018}, while others rely on 3D images \cite{Chen-NeuroImage-2018,Kamnitsas-MIA-2017,Gibson-TMI-2018,Luo-MIA-2021,Zhao-TMI-2022,Chen-MIA-2022,Wolz-TMI-2013,Lyksborg-SCIA-2015, Baldeon-NN-2020,Seo-TMI-2020,Jin-MIA-2021}. The works using 2D images as input naturally produce 2D slices as output, while the works using entire 3D volumes as input produce 3D volumes as output.

Perhaps the most popular architecture for 2D segmentation is U-Net~\cite{Ronneberger-MICCAI-2015}. U-Net is a fully convolutional (conv) network designed for medical image segmentation. The architecture follows a ``U'' shape and is composed of a contracting and an expansive path. Each step of the expansive path is composed of an upsampling operation, a convolution layer which halves the number of feature maps, and a concatenation with the corresponding cropped feature maps from the contracting path. Seo et al.~\cite{Seo-TMI-2020} proposed the mU-Net model, a modified version of the U-Net architecture. mU-Net~\cite{Seo-TMI-2020} adds a residual path to the deconvolution operations, and an additional convolutional layer to the skip connections in order to extract high-level global features of small objects.

Chen et al.~\cite{Chen-NeuroImage-2018} proposed the voxel-wise residual network (VoxResNet), a 3D CNN formed of 25 layers with residual connections. Multimodal and multi-level contextual information is introduced into the VoxResNet model. The multimodal information is added by concatenating multimodal data before giving it as input to the model. To improve the 3D segmentation performance of brain lesions, Kamnitsas et al.~\cite{Kamnitsas-MIA-2017} employed a 3D CNN comprising 11 layers with parallel convolutional pathways for multi-scale processing. Rather than modifying the layers of their architecture, Zhao et al.~\cite{Zhao-TMI-2022} inserted a lesion-related spatial attention mechanism into the network.

In order to help physicians obtain better segmentation results, Luo et al.~\cite{Luo-MIA-2021} proposed interactive segmentation to further improve the performance of CNN models, even to unseen objects.

Closer to our study, the work of Gibson et al.~\cite{Gibson-TMI-2018} shares the same target task, being focused on multi-organ abdominal segmentation. Gibson et al.~\cite{Gibson-TMI-2018} presented a registration-free approach based on Dense V-Networks for multi-organ abdominal segmentation of 3D images. They also proposed a batch-wise spatial dropout to lower the memory usage and processing time of dropout.

Different from the aforementioned works, which are trained in a fully-supervised learning setting, there are several works~\cite{Zhou-ICCV-2019,Chen-CVPR-2022} proposing weakly-supervised learning frameworks. Zhou et al.~\cite{Zhou-ICCV-2019} found that data sets having only one organ annotated as the positive class, leaving the other organs as part of the background, attain misleading results in multi-organ segmentation, since the background class contains many organs. In order to alleviate this problem, Zhou et al.~\cite{Zhou-ICCV-2019} proposed a prior-aware neural network, incorporating anatomical priors on abdominal organ sizes into the training objective.
 
Similar to our approach proposing an ensemble of multiple networks to improve the segmentation results, Lyksborg ~\cite{Lyksborg-SCIA-2015} proposed to use a model for each of the axial, sagittal and coronal planes, fusing the corresponding segmentations into a single 3D segmentation. Baldeon et al.~\cite{Baldeon-NN-2020} proposed AdaEn-Net, an ensemble of networks that boosts the segmentation performance. AdaEn-Net~\cite{Baldeon-NN-2020} firstly employs an ensemble of 2D and 3D models to predict the output segmentation. Then, it trains the 2D-3D ensemble on $k$-folds, obtaining $k$ models. The final segmentation mask is the average of the $k$ models forming the final ensemble.

Different from previous works, such as~\cite{Lyksborg-SCIA-2015,Baldeon-NN-2020}, which directly combined models into ensembles without taking into account their output correlation, we propose a novel ensemble creation algorithm which promotes the diversity among the models comprising the ensemble.

\section{Method}

\subsection{Neural Architectures}

To address our medical image segmentation task, we employ the well known U-Net architecture~\cite{Ronneberger-MICCAI-2015}. The U-Net architecture is a fully convolutional network that belongs to the family of encoder-decoder neural networks. In the encoding part, the spatial information is downsampled through convolution and pooling operations. In the decoding part, the spatial information is upsampled back to the original size via convolution transpose. High-resolution features from the encoder are passed through skip connections and concatenated to the corresponding features from the decoder, thus infusing high-resolution information into the decoder. The introduction of skip connections gives the network its ``U'' shape. We further present our changes to the U-Net model, leading to a total of nine distinct model variants forming the basis of our ensemble.

\begin{figure*}[!th]
\includegraphics[width=0.8\linewidth]{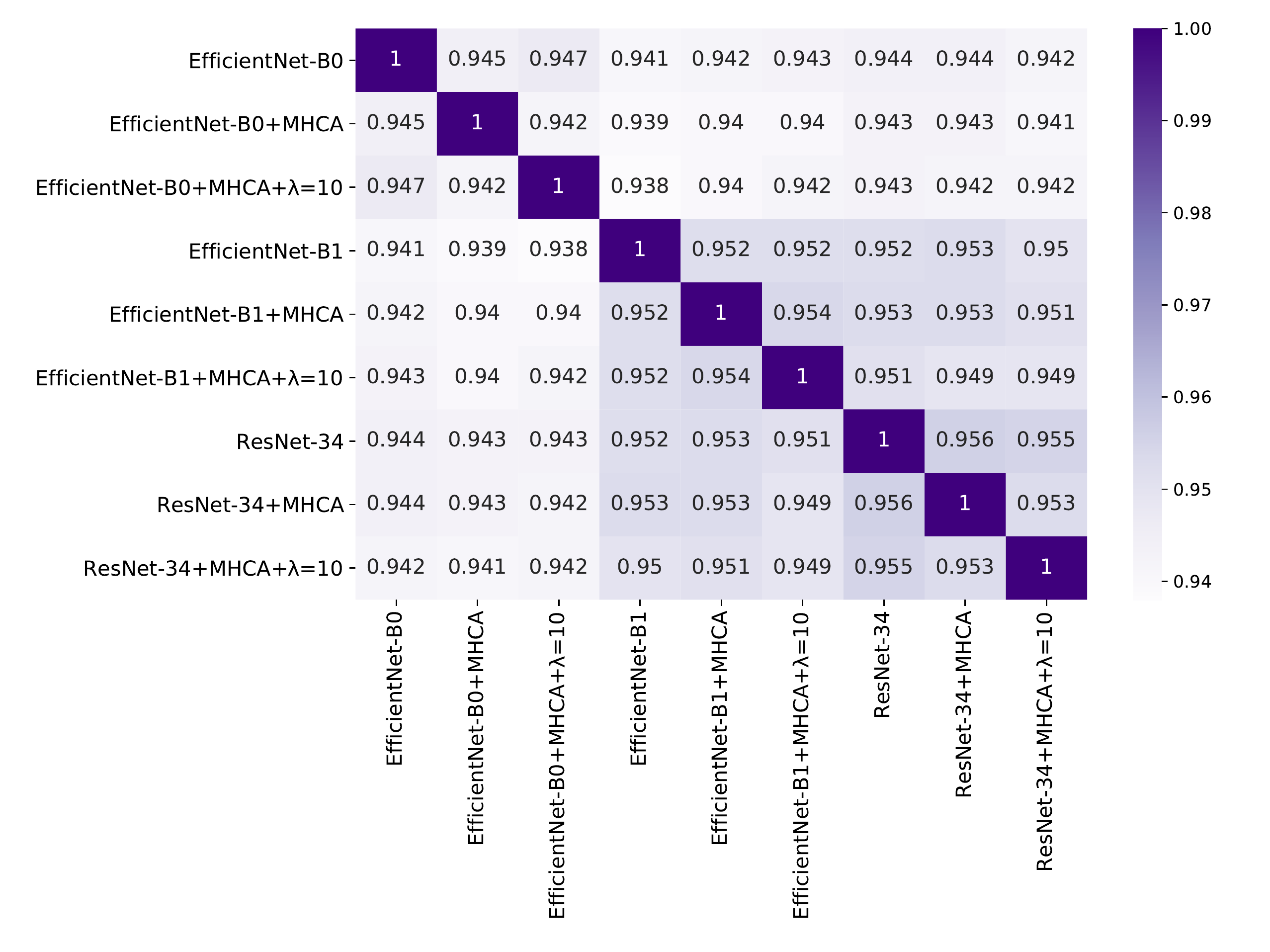}
\vspace{-0.4cm}
\caption{Correlation matrix based on paired Dice coefficients between the individual U-Net models included in our study.\label{fig_corr_mat}}
\end{figure*}

\subsubsection{Backbone Variations}

To build an ensemble of a diverse set of models, we first introduce variations in terms of the backbone architecture. Therefore, we try the following three encoder architectures: ResNet-34 \cite{He-CVPR-2016}, EfficientNet-B0 \cite{Tan-ICML-2019}, and EfficientNet-B1 \cite{Tan-ICML-2019}. We choose ResNet-34 due to its fairly good trade-off between running time and accuracy level. The reason behind adding EfficientNet-B0 and EfficientNet-B1 into our study is the superior performance levels of these models compared to ResNet-34. 

The residual network (ResNet) architecture was proposed by He et al.~\cite{He-CVPR-2016}. ResNet models are composed of residual blocks. A residual block consists of a few stacked conv layers and a skip connection from the first layer to the last layer of the block. Skip connections allow the training of very deep neural networks, alleviating the vanishing gradient problem. He et al.~\cite{He-CVPR-2016} proposed five ResNet variants of different depth, namely ResNet-18, ResNet-34, ResNet-50, ResNet-101 and ResNet-152. Among these, we select ResNet-34 to serve as backbone for some of our U-Net models. 

The EfficientNet architecture was introduced by Tan et al.~\cite{Tan-ICML-2019} to efficiently scale convolutional neural networks. 
Tan et al.~\cite{Tan-ICML-2019} demonstrated that, in order to obtain better performance under a certain computational budget, all three components of the network, namely the depth, the width and the resolution, should be uniformly scaled. Based on this finding, the EfficientNet family of models (from B1 to B7) was created starting from the EfficientNet-B0 architecture and scaling the depth, width and resolution of each network version by a certain factor. EfficientNets generally obtain better performance than the previous state-of-the-art convolutional networks under the same computational resources.

We employ the same decoder architecture, regardless of the encoder type. More specifically, the decoder is formed of five successive convolutional blocks with the following number of filters: $512$, $256$, $128$, $64$ and $64$. Inside the decoder, we use nearest neighbor interpolation as the upsampling operation.

\subsubsection{Integrating Attention}

In order to further increase the variability of the studied models, we add a newly introduced multi-head convolutional attention (MHCA)~\cite{Georgescu-WACV-2023} mechanism. 
MHCA performs both channel and spatial attention by applying multiple convolutional attention heads. Each convolutional head has a distinct receptive field size corresponding to a particular reduction rate for the spatial attention. At the same time, the number of filters is used to control the channel reduction rate. The MHCA~\cite{Georgescu-WACV-2023} block can be introduced at any layer of any neural architecture.

We underline that the MHCA~\cite{Georgescu-WACV-2023} block was originally introduced for medical image super-resolution. We opt for introducing the MHCA module instead of other popular attention mechanisms, e.g.~Squeeze-and-Excitation~\cite{Hu-CVPR-2018} or CBAM~\cite{Woo-ECCV-2018}, because the former is specifically tailored for medical images. 

\subsubsection{Loss Variations}
 
When performing semantic segmentation, especially in medical images, there is a high risk of class imbalance due to the prevalence of the background class. In order to alleviate the class imbalance problem, we propose a modification to the loss function that assigns higher weights to the positive classes.

The loss function that we employ to optimize the models is:
\begin{equation}\label{eq_total_loss} 
\mathcal{L}_{\scriptsize{\mbox{total}}}(Y, \hat{Y}) = 0.5 \cdot \mathcal{L}_{\scriptsize{\mbox{BCE}}}(Y, \hat{Y}) + 0.5 \cdot \mathcal{L}_{\scriptsize{\mbox{Tversky}}}(Y, \hat{Y}).
\end{equation}

The binary cross entropy (BCE) loss function is defined as:
\begin{equation} 
\mathcal{L}_{\scriptsize{\mbox{BCE}}}(Y, \hat{Y}) = -(\lambda \cdot Y \cdot log(\sigma(\hat{Y})) + (1 - Y) \cdot log(1 - \sigma(\hat{Y}))),
\label{bce}
\end{equation}
where $\lambda = 1$. The Tversky loss function~\cite{Salehi-MLMI-2017} is defined as follows:
\begin{equation} 
\mathcal{L}_{\scriptsize{\mbox{Tversky}}}(Y, \hat{Y})\!=\!\frac{\sum_{i=1}^m{\hat{y}_{0,i}\!\cdot\! y_{0,i}}}{\sum_{i=1}^m{\hat{y}_{0,i}\!\cdot\!y_{0,i}}\!+\! \alpha\!\cdot\!\sum_{i=1}^m{\hat{y}_{0,i}\!\cdot\!y_{1,i}}\!+\! \beta\!\cdot\!\sum_{i=1}^m{\hat{y}_{1,i}\!\cdot\! y_{0,i}}},
\end{equation}
where $Y$ is the ground-truth label, $\hat{Y}$ is the predicted value, $y_{0,i}$ is a ground-truth background (negative) voxel, $y_{1,i}$ is a ground-truth organ (positive) voxel, $\hat{y}_{0,i}$ is a predicted background voxel, $\hat{y}_{1,i}$ is a predicted organ voxel, and $m$ is the number of voxels. We use the default values of $\alpha=0.5$ and $\beta=0.5$.  

We consider two configurations for the loss defined in Eq.~\eqref{bce}, one where $\lambda=1$ and another where $\lambda=10$. When using the default value of $1$ for $\lambda$, the positive and the negative examples have an equal influence on the loss function. This is known as the standard BCE loss. By assigning a higher weight ($\lambda=10$) to the positive classes, the network will receive a higher penalty when it classifies a non-background pixel as background. In other words, our positively-biased BCE loss (based on $\lambda=10$) reduces the number of false negative pixels.

\subsection{Diversity-Promoting Ensemble}

We consider the case when a processing budget is imposed for the ensemble, specifically by limiting the number of models that can enter the ensemble. Perhaps the most straightforward solution to create an ensemble with a limited number of models is to simply select the models with the highest performance from a pool of models. We consider this ensemble creation strategy as a competitive and representative baseline for our own strategy.

We conjecture that in order to have a more accurate ensemble, the output of individual models forming the ensemble should not be correlated. If the models in the ensemble are correlated, the aggregated output will likely not be very different from the output of a single model. As mentioned earlier, our first step is to train a diverse set of U-Net models with the scope of creating a strong ensemble, promoting diversity of the included models. More specifically, we propose a new strategy to create an ensemble by taking into account the correlation between the outputs of individual networks. We provide empirical evidence showing that promoting the decorrelation, i.e.~reducing the correlation, of the outputs of multiple networks can lead to more accurate ensembles. 

As an upper bound for the ensembles with a limited budget, we lift any budget limitation, considering an ensemble that puts all the available models together.

All ensembles, including our own, are based on predicting the final output by computing the average of the softmax activations of the individual models. This is known in literature as soft plurality voting. 


We further present our diversity-promoting ensemble (DiPE) creation strategy, which is based on an algorithm that relies on the correlation matrix between pairs of models.

\begin{algorithm}[t]
\caption{Diversity-Promoting Ensemble Creation Algorithm\label{alg_ensemble}}
\small{
\textbf{Input}: 

$n$ -- the total number of models.

$k, 1 \le k \le n$ -- the maximum number of models for the ensemble.

$\mathcal{M} = \{ M_1, M_2, ..., M_n\}$ -- the set of segmentation models.

\BlankLine
\textbf{Initialization}:

$C \leftarrow$ compute the correlation matrix between the $n$ models via Eq.~\eqref{eq_corr}.

$D = \{d_1, d_2, ...., d_n \} \leftarrow$ compute the Dice coefficients of the $n$ models with respect to the ground-truth validation labels.

\textbf{Computation}:

$s \leftarrow \argmax_i(d_i)$ (get index of the best individual model)

$\mathcal{E} \leftarrow \{M_{s}\}$ (select the best individual model)

\For{$\;j \in \{1,2,..., k - 1 \}$}
{   
    $s \leftarrow 0$ (initialize index of model to be added)
    
    $c \leftarrow \infty$ (initialize correlation of model to be added)
    
    $p \leftarrow 0$ (initialize performance of model to be added)
    
    \For{$\;i \in \{1,2,..., n\}$}
    {
        \If{$M_i \in \mathcal{M} \setminus \mathcal{E}$}
        {
            $C_{i,\mathcal{E}} \leftarrow$ compute the average correlation via Eq.~\eqref{eq_avg_corr}.
            
            \If{$C_{i,\mathcal{E}} < c\;\mbox{\textbf{or}}\; \left(C_{i,\mathcal{E}} = c \;\mbox{\textbf{and}}\; d_i > p \right)$}
            {
                $s \leftarrow i$ 
    
                $c \leftarrow C_{i,\mathcal{E}}$
    
                $p \leftarrow d_i$
            }
        }
    }
    
    $\mathcal{E} \leftarrow \mathcal{E} \cup \{M_s\}$
}

\BlankLine
\textbf{Output}: 

$\mathcal{E}$ -- the ensemble consisting of $k$ models.
}
\end{algorithm}

\subsubsection{Correlation Matrix}

Before applying our algorithm, we need to compute the pairwise correlation matrix between the outputs of different models. To determine the correlation score between two models $M_i$ and $M_j$, we employ the Dice coefficient computed between the output segmentations provided by the respective networks for the entire validation set $\mathcal{X}=\{X_1, X_2, ..., X_t\}$, where $t$ is the number of validation samples. We underline that the Dice score is a commonly used measure to evaluate how close the output of a model is to the ground-truth segmentation. In our case, we propose to simply assume that the output of one model is regarded as the ground-truth and the output of the other model is regarded as the prediction. Since the Dice coefficient is symmetric, it does not matter which model is regarded as ground-truth. For some medical image $X_r \in \mathcal{X}$, let $\hat{Y}_{i,r}$ and $\hat{Y}_{j,r}$ denote the segmentations given as output by the models $M_i$ and $M_j$, i.e.:
\begin{equation}
\hat{Y}_{i,r} = M_i(X_r), \hat{Y}_{j,r} = M_j(X_r).
\end{equation}

With the above notations, the Dice coefficient between $\hat{Y}_{i,r}$ and $\hat{Y}_{j,r}$ is defined as:
\begin{equation}\label{eq_Dice} 
\mbox{Dice}(\hat{Y}_{i,r}, \hat{Y}_{j,r}) = \frac{2 \cdot \lvert \hat{Y}_{i,r} \cap \hat{Y}_{j,r} \rvert}{ \lvert \hat{Y}_{i,r} \rvert +  \lvert \hat{Y}_{j,r} \rvert}.
\end{equation}

Next, we define the correlation $C_{i,j}$ between $M_i$ and $M_j$ as follows:
\begin{equation}\label{eq_corr} 
C_{i,j} = \frac{1}{t} \sum_{r=1}^t \mbox{Dice}(\hat{Y}_{i,r}, \hat{Y}_{j,r}).
\end{equation}

By applying Eq.~\eqref{eq_corr} to all possible model pairs formed from $n$ models, we obtain a symmetric correlation matrix $C$ of size $n\times n$. We illustrate an example of a correlation matrix in Figure~\ref{fig_corr_mat}. The highest correlation is naturally obtained between a model $M_i$ and itself. Hence, the values on the diagonal of $C$ are equal to $1$.

\subsubsection{Ensemble Construction Algorithm}

Given a set of $n$ models $\mathcal{M} = \{ M_1, M_2, ..., M_n\}$, and the Dice coefficients $D = \{d_1, d_2, ...., d_n \}$ of the $n$ models with respect to the ground-truth validation labels, we start building the ensemble $\mathcal{E}$ by choosing the best available model. Then, in order to add a new model from $\mathcal{M} \setminus \mathcal{E}$ to $\mathcal{E}$, we select the best model which is also the least correlated with the models currently added to $\mathcal{E}$. For each remaining model $M_i \in \mathcal{M} \setminus \mathcal{E}$, we determine the correlation score between $M_i$ and $\mathcal{E}$ by taking the average of the correlation scores between $M_i$ and the models from $\mathcal{E}$. In addition, we sum up the Dice error $1-d_i$ of the model $M_i$ to each correlation $C_{i,j}$, thus giving equal importance to the performance level of $M_i$ and the correlation score between $M_i$ and $M_j$. Finally, the average score for adding $M_i$ to $\mathcal{E}$ is computed as follows: 
\begin{equation}\label{eq_avg_corr} 
C_{i,\mathcal{E}} = \avg\left((1 - d_i) + C_{i,j}\right), \forall j\; \mbox{such that} \; M_j \in \mathcal{E}.
\end{equation}

If there are multiple models with an average score equal to the minimum average score at a certain step, we select the best performing model among them and add it to $\mathcal{E}$. We perform this operation for $k$ steps, where $k$ ($1 \le k \le n$) is the maximum number of models to be added to the ensemble, i.e.~$k$ represents the budget limitation for the ensemble. The whole process is formally described in Algorithm~\ref{alg_ensemble}. 

We emphasize that our ensemble creation strategy does not require any additional learning steps. Moreover, it does not add any running time overhead during inference. Hence, its computational cost at test time is similar to that of the baseline ensemble creation strategy.


\section{Experiments}

\begin{figure*}
\includegraphics[width=0.95\linewidth]{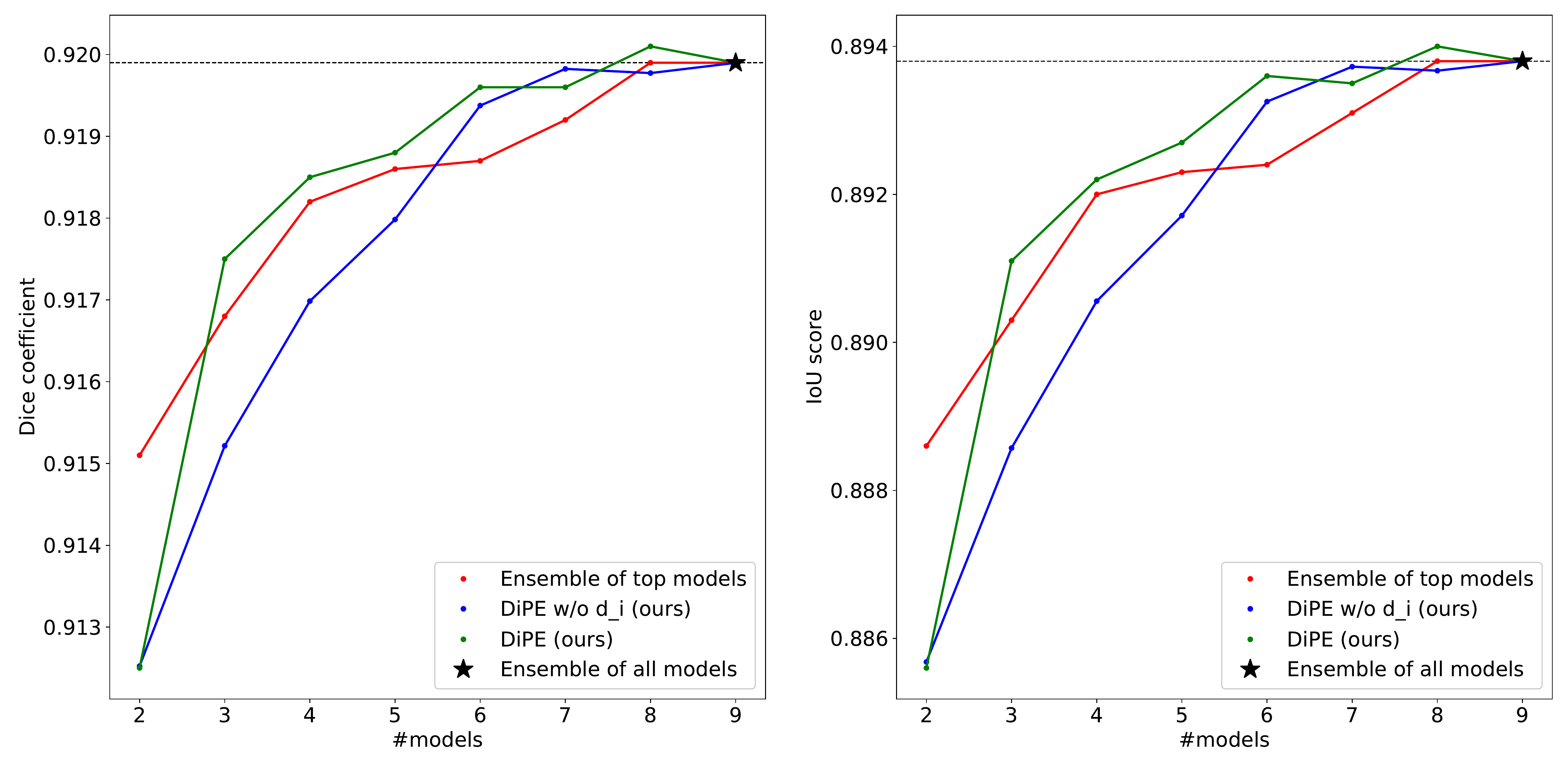}
\vspace{-0.35cm}
\caption{Comparing the baseline ensemble formed of the top $\mathbf{k}$ models (red) with DiPE (green) and ablated DiPE (blue), for $\mathbf{k \in \{2,3,...,9 \}}$. The dashed horizontal line shows the results when including all nine U-Nets into the ensemble (no budget limitation on the number of models). The ablated version of DiPE is obtained by removing the Dice error ($\mathbf{1-d_i}$) with respect to the ground-truth validation labels from Eq.~\eqref{eq_avg_corr}. Best viewed in color.\label{fig_ensemble_9}}
\end{figure*} 

\begin{figure*}[t] 
\includegraphics[width=0.78\linewidth]{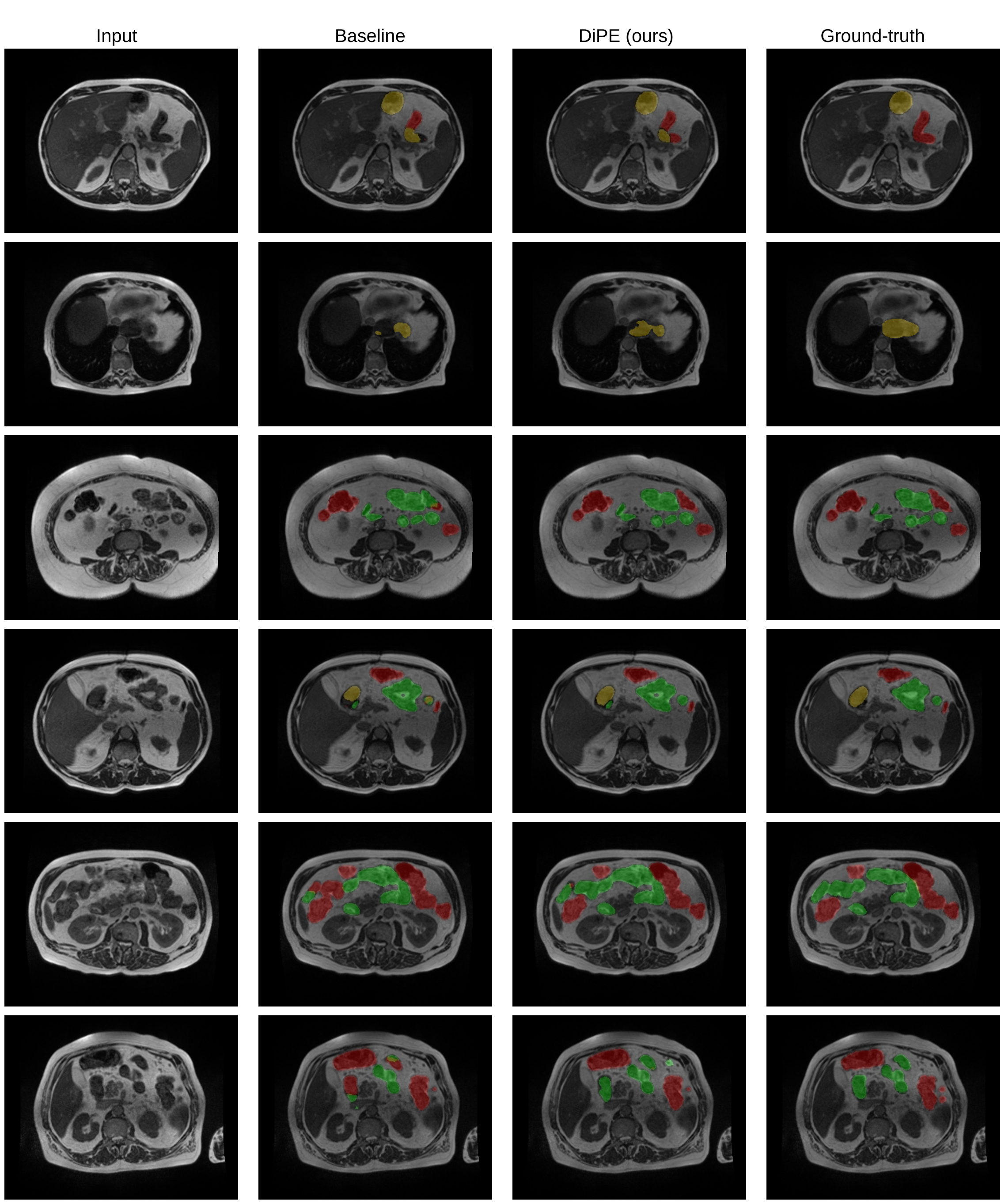}
\vspace{-0.25cm}
\caption{Comparing the quality of a set of segmentation results given by the baseline ensemble formed of the top $\mathbf{k}$ models (second column) with the segmentation results produced by DiPE (third column). The original inputs (first column) and the ground-truth segmentation masks (fourth column) are included for reference. Best viewed in color.\label{fig_viz}}
\end{figure*}   

\subsection{Data Set}

The UW-Madison GI Tract Image Segmentation data set \cite{Lee-Kaggle-2022} contains MRI scans of the abdominal area. The data set comprises a total of $38,496$ slices. There are three annotated organs, namely the stomach, the small bowel and the large bowel. Only $21,906$ out of $38,496$ slices have annotated organs, the rest of the slices showing abdominal regions where these three organs are not seen, thus being completely annotated as background. The size of each slice varies between $234\times234$ and $384\times384$ pixels. We split the data set into $80\%$ for training and $20\%$ for testing. We keep $10\%$ of the training data for validation purposes.

\subsection{Evaluation Metrics}

In order to evaluate the models, we employ two evaluation metrics. Both metrics measure the overlap between the ground-truth and the predicted segmentations. The first measure is the Intersection over Union (IoU). The IoU score between a ground-truth segmentation mask $Y$ and a predicted mask $\hat{Y}$ is defined as:
\begin{equation} 
\mbox{IoU}(Y, \hat{Y}) = \frac{\lvert Y \cap \hat{Y} \rvert}{\lvert Y \cup \hat{Y} \rvert}.
\end{equation}

The second evaluation metric is the Dice coefficient between the ground-truth segmentation and the predicted mask, defined as:
\begin{equation} 
\mbox{Dice}(Y, \hat{Y}) = \frac{2  \cdot \lvert Y \cap \hat{Y} \rvert}{ \lvert Y \rvert +  \lvert \hat{Y} \rvert}.
\end{equation}

\subsection{Implementation Details}

We implement the models in PyTorch. Before training the models, we initialize the weights of the backbones with the values pre-trained on ImageNet~\cite{Russakovsky-IJCV-2015}. We set the input size to $320\times384$, regardless of the chosen encoder. We train each network for $20$ epochs, setting the batch size to $32$ and the learning rate to $10^{-3}$.  We optimize the networks using the loss defined in Eq.~\eqref{eq_total_loss}, employing the Adam optimizer~\cite{Kingma-ICLR-2014} to update the parameters of the models.

When integrating the MHCA modules~\cite{Georgescu-WACV-2023}, we keep the optimal hyperparameters specified by the authors, namely the reduction ratio equal to $0.5$ and the number of heads set to $3$. We add MHCA to the skip connections of the U-Net model. Based on a preliminary evaluation conducted on the validation set, we decided to add MHCA only to the last two skip connection layers. This can be motivated by the fact that the early skip connection feature maps have very few channels.
 
\subsection{Results}

\begin{table}
\caption{Dice and IoU scores obtained by individual U-Net model variants based on different backbones (ResNet-34, EfficientNet-B0, EfficientNet-B1), with and without MHCA modules \cite{Georgescu-WACV-2023}, and alternating between the standard BCE loss and our positively-biased BCE loss with $\mathbf{\boldsymbol{\lambda}=10}$.\vspace{-0.2cm}} \label{tab_main_results}
\begin{tabular}{ccccc}
\toprule
\textbf{Backbone} & \textbf{MHCA~\cite{Georgescu-WACV-2023}}  & $\mathbf{\boldsymbol{\lambda}=10}$ & \textbf{Dice}	& \textbf{IoU}\\

\midrule
\multirow{3}{*}{ResNet-34}          & \xmark & \xmark &  $0.9051$			& $0.8774$\\
		                            & \cmark & \xmark &  $0.9045$			& $0.8769$\\
        		                    & \cmark & \cmark & $0.9054$			& $0.8780$\\
\midrule
\multirow{3}{*}{EfficientNet-B0}          & \xmark & \xmark &  $0.9111$			& $0.8836$\\
		                                   & \cmark & \xmark &  $0.9087$			& $0.8814$\\
        		                           & \cmark & \cmark &  $0.9103$			& $0.8830$\\
\midrule
\multirow{3}{*}{EfficientNet-B1}          & \xmark & \xmark &  $0.9110$			& $0.8843$\\
		                                   & \cmark & \xmark &  $0.9112$			& $0.8844$\\
        		                           & \cmark & \cmark &  $0.9130$			& $0.8860$\\        		                           
\bottomrule
\end{tabular}
\end{table}

\begin{table} 
\caption{Comparing the baseline ensemble formed of the top $\mathbf{k}$ models with our diversity-promoting ensemble (DiPE), for $\mathbf{k \in \{2,3,...,9 \}}$. The last row shows the results when including all nine U-Nets into the ensemble (no budget limitation on the number of models). When $\mathbf{k=9}$, DiPE becomes equivalent with the ensemble based on all models. Best overall scores are highlighted in bold.\vspace{-0.2cm}}\label{tab_ensemble}
\begin{tabular}{ccc|cc}
\toprule
 \textbf{\#models} &  \multicolumn{2}{c|}{\textbf{Baseline}}  & \multicolumn{2}{|c}{\textbf{DiPE (ours)}}    \\
$\mathbf{(k)}$                   & \textbf{Dice}  & \textbf{IoU} & \textbf{Dice}  & \textbf{IoU}  \\

\midrule
                                     2 &  $0.9151$	    	& $0.8886$ &    $0.9125$			& $0.8856$\\  
                                     3 &  $0.9168$			& $0.8903$ &    $0.9175$			& $0.8911$\\  
                                     4 &  $0.9182$			& $0.8920$ &    $0.9185$			& $0.8922$\\  
                                     5 &  $0.9186$			& $0.8923$ &    $0.9188$			& $0.8927$\\  
                                     6 &  $0.9187$			& $0.8924$ &   $0.9196$ 		    & $0.8936$ \\  
                                     7 &  $0.9192$			& $0.8931$ &   $0.9196$			& $0.8935$\\  
                                     8 &  $0.9199$			& $0.8938$ &   $\mathbf{0.9201}$			& $\mathbf{0.8940}$\\ 
\midrule 
All (9)                                      &  $0.9199$			& $0.8938$ &    $0.9199$			& $0.8938$\\  
\bottomrule
\end{tabular}
\end{table}

\subsubsection{Quantitative Results of Individual U-Net Models}
We show the results obtained by individual models on the UW-Madison GI Tract Image Segmentation data set in Table~\ref{tab_main_results}. We observe that the best performance of $0.9130$ in terms of the Dice coefficient is obtained when the EfficientNet-B1 architecture is used as backbone for the encoder. When using the ResNet-34 encoder, the Dice coefficient drops to $0.9051$, showing that the encoder capacity can have a significant impact on the results. 

When adding the MHCA modules, only the model based on the EfficientNet-B1 backbone obtains an improvement in terms of the IoU and Dice scores. For the other two models, namely EfficientNet-B0 and ResNet-34, the performance drops by a small margin. When introducing the positively-biased BCE loss (with $\lambda=10$ in Eq.~\eqref{bce}), we observe considerable performance gains for the ResNet-34 and EfficientNet-B1 backbones.

\subsubsection{Quantitative Results of Ensembles}

Starting from the nine individual U-Net models, we evaluate two ensemble creation strategies which have a restricted computational budget. The budget limitation is introduced in terms of the number of models forming each ensemble. In Table~\ref{tab_ensemble}, we present the results of the baseline ensemble creation strategy (based on selecting the models with the best performance) versus the results of our DiPE strategy. We observe that our method generally surpasses the baseline method, the only exception being when the number of models $k$ is equal to $2$. We also note that DiPE obtains slightly better results when using only $k=8$ models in the ensemble than the baseline that combines all  models ($0.9201$ versus $0.9199$ in terms of the Dice coefficient). We can also visualize the performance of our approach versus the performance of the baseline method in Figure~\ref{fig_ensemble_9}. The figure clearly shows that DiPE produces superior results, for all ${k \in \{3,...,8 \}}$.

\begin{table} 
\caption{Ablation results obtained by removing the Dice error ($\mathbf{1-d_i}$) with respect to the ground-truth validation labels from Eq.~\eqref{eq_avg_corr}.\vspace{-0.2cm}}\label{tab_ablation}
\begin{tabular}{ccc|cc}
\toprule
 \textbf{\#models} &  \multicolumn{2}{c|}{\textbf{DiPE w/o $\mathbf{d_i}$ (ours)}}  & \multicolumn{2}{|c}{\textbf{DiPE (ours)}}    \\
$\mathbf{(k)}$                   & \textbf{Dice}  & \textbf{IoU} & \textbf{Dice}  & \textbf{IoU}  \\

\midrule
                                     2 &     $0.9125$ &  $0.8856$ &    $0.9125$			& $0.8856$\\   
                                     3 &     $0.9152$ &  $0.8886$ &    $0.9175$			& $0.8911$\\   
                                     4 &     $0.9170$ &  $0.8906$ &    $0.9185$			& $0.8922$\\  
                                     5 &     $0.9180$ &  $0.8917$ &    $0.9188$			& $0.8927$ \\  
                                     6 &     $0.9194$ &  $0.8933$ &   $0.9196$ 		    & $0.8936$  \\  
                                     7 &     $0.9198$ &  $0.8937$ &   $0.9196$ 		    & $0.8935$ \\  
                                     8 &     $0.9198$ &  $0.8937$ &   $0.9201$			& $0.8940$\\  
\bottomrule
\end{tabular}
\end{table}

\subsubsection{Ablation Results}
We perform an ablation study to prove the necessity of adding the Dice error of the model when computing the average correlation score in Eq.~\eqref{eq_avg_corr}. We present the corresponding results in Table~\ref{tab_ablation}. If we do not take into account the performance of the models with respect to the ground-truth when creating the ensemble, the performance levels of the ensembles with various budgets drop considerably, e.g.~from $0.9188$ to $0.9180$ in terms of the Dice coefficient when using $k=5$ models. This can be explained by the fact that a model which is not correlated with the models already included in the ensemble, might also be less correlated with the ground-truth. In other words, its low correlation score with the previously included models is caused by the poor performance level of the respective model. Hence, we observe that the best strategy when creating an ensemble is to take both aspects into account, namely the performance of the new model and its correlation with the models comprising the current ensemble.

\subsubsection{Qualitative Results of Ensembles}
In Figure~\ref{fig_viz}, we illustrate some qualitative results to compare the baseline ensemble strategy with our DiPE strategy. We show the segmentation masks of the ensembles consisting of $k=5$ models. We notice that DiPE creates an ensemble which produces better segmentation masks than the baseline strategy. From a medical point of view, the comparison between our DiPE method and the ground-truth (produced by physicians specialized in radiology) exhibits very few differences. For example, we particularly observe that DiPE obtains a segmentation of the stomach (second row of Figure~\ref{fig_viz}) that is much closer to the ground-truth segmentation than the baseline method. In fact, the first two rows exemplify the ability of DiPE to identify and differentiate between the stomach and the transverse colon with a better accuracy than the baseline strategy, producing a result close to what the radiologists have annotated. In general, the stomach region is well identified by both methods, with a slight upper hand for our technique.
Furthermore, we underline that our method especially excels on the tissues surrounded by larger amounts of fat, such as the small bowel and the mesentery, as seen in the bottom images. We consider that the ability of DiPE to distinguish between the small bowel and the colon is outstanding, being an almost one-to-one match with the ground-truth. Remarkably, the example shown on the fourth row illustrates that our technique does not mistake the loop of small bowel on the right (or the patient's left) with gastric structures.
In summary, our new method offers significantly better results in comparison with the baseline method, but we believe there is still room for improvement in order to reach the full precision of the experimented human eye.


\section{Conclusion}

In this work, we studied the task of medical image segmentation, specifically focusing our experiments on the segmentation of the stomach, the small bowel and the large bowel. Towards improving the segmentation accuracy, we proposed a bottom-up diversity-promoting ensemble creation strategy that gives equal importance to the accuracy level and the correlation with other models, when adding a new model to the ensemble. Our ablation results showed that using solely the correlation with other models is suboptimal.

In future work, we aim to extend our empirical analysis to more segmentation tasks from the medical domain. We also aim to replace plurality voting with a meta-learner, which could further improve performance.

\section*{Acknowledgements}

The research leading to these results has received funding from the NO Grants 2014-2021, under project ELO-Hyp contract no. 24/2020. 

\bibliographystyle{ACM-Reference-Format}
\bibliography{references} 

\end{document}